\documentclass[]{aa}
\usepackage{graphicx}

\begin{document}

\title{Integration of the atmospheric fluctuations in a dual-field 
optical interferometer: the short exposure regime }

\author{G\'erard Daigne\inst{1}
  \and Jean-Fran\c{c}ois Lestrade\inst{2}}
 
\offprints{G. Daigne}

\institute{Observatoire Aquitain des Sciences de l'Univers, B.P.89, 
  F--33270 Floirac, France \\
  email: daigne@obs.u-bordeaux1.fr
 \and 
  Observatoire de Paris-CNRS, 77 Ave. Denfert-Rochereau, 
  F--75014 Paris, France \\
   email: lestrade@obspm.fr
   }

\date{Received 9 Oct 2002 /Accepted 18 Apr 2003}
\titlerunning{Dual-field optical interferometer}
\authorrunning{Daigne \& Lestrade}

\abstract{
Spatial phase-referencing in dual-field optical interferometry
is reconsidered. Our analysis is based on 
the 2-sample variance of the differential phase between target and reference star.
We show that averaging over time of the atmospheric effects depends on this 
2-sample phase variance (Allan variance) rather than on the true variance. 
The proper expression for fringe smearing beyond the isoplanatic angle 
is derived.
With simulations of atmospheric effects, based on a Paranal turbulence model,  
we show how the performances of a dual-field optical interferometer can be 
evaluated in a diagram 'separation angle' versus 'magnitude of faint object'.
In this diagram, a domain with short exposure is found to be most useful for 
interferometry, with about the same magnitude limits in the $H$ and $K$ bands. 
With star counts from a Galaxy model, we evaluate the sky coverage for 
differential astrometry and detection of exoplanets, i.e. likelihood of 
faint reference stars in the vicinity of a bright target. With the 
2mass survey, we evaluate sky coverage for phase-referencing, i.e. 
avaibility of a bright enough star for main delay tracking in the vicinity 
of any target direction.
}
\maketitle

%%-----------------------------
\section{Introduction}

Phase-referencing in ground-based optical interferometry has been proposed
for faint object imaging, for narrow-angle astrometry and for improved precisions 
in visibility measurements (\cite{mozurk88}; \cite{cola92}; \cite{quirren94}). 
Phase-referencing is the interferometer counterpart of Adaptive Optics (AO)
developed for single-pupils. 
Inhomogeneities in the atmosphere cause the optical path difference (OPD) between 
two telescopes to fluctuate over a range of time scales, from a few 
milliseconds upwards, with variance rms of about 10 $\mu$m. These fluctuations 
can be measured continuously on a bright star and used to track a faint star  
in the immediate surrounding. Then faint objects can 
be observed and small visibility amplitude measured with sufficient Signal/Noise ratio. 
Furthermore, differential OPD fluctuations ($\Delta$OPD) between target 
and reference stars should average out over long integrations, leading to 
differential phase structure for faint objects partially resolved, 
or to high precision astrometry for unresolved sources. 

It has been noticed that $\Delta$OPD residuals over long integrations 
are proportional to the measured angle (\cite{shaocol}), 
so that high precision narrow-angle astrometry needs to be performed with  
reference sources in the immediate surrounding of the source target, 
i.e. within the isoplanatic patch. 
Angular anisoplanatism in optical interferometry has been recently investigated 
(\cite{espos2000}). The authors define an ``isopistonic'' angle where 
the interferometric $\Delta$OPD (rms value) remains smaller than a tenth of 
the observed wavelength. Their analysis is based on the true variance of 
differential piston. A detailed analysis of the 
$\Delta$OPD spectrum has been performed by \cite{darcio} who considers also 
its jitter on short exposures, but does not investigate its potential use 
in phase-referencing. 

With the condition ``differential phase rms smaller than 1 radian'', 
the half-cone angle of Spatial Phase-Referencing (SPR) is found to be the isoplanatic 
angle for AO (\cite{fried82}), divided by ($2^{3/5}$) due to 
the two entrance pupils of a long baseline interferometer with uncorrelated 
phase corrugations (\cite{cola92}). 
Taking into account the aperture size, differential phase excursion remains 
bounded even on separation angles exceeding slightly the AO isoplanatic angle, 
and SPR with long integrations is possible. 
On larger angular separations, random walk of 
differential phase fluctuations prevents long term averaging. The measurement 
of differential piston variations is still possible, but with short integrations, 
shorter than some coherence time for $\Delta$OPD fluctuations. Sensitivity is 
reduced as compared to small angles, but the chance of finding a bright enough 
star for phase-referencing is increased. Altogether, SPR in optical interferometry 
has to be investigated in the whole telescope Field-of-View, that is within 
a few arcminutes. 

An optimization of SPR requires a detailed analysis of the temporal 
behaviour of differential phase. 
Our analysis is based on the 2-sample (or Allan) variance of time series 
of the measured differential phase. This quantity 
tells us how $\Delta$OPD variations can be recovered before being averaged. 
Ultimately, the true $\Delta$OPD variance will give the achievable precision 
on long integrations. 

Here we disentangle the problem between differential piston fluctuations and 
single-pupil wavefront distortion, that is between the first mode and 
higher modes in the Zernike expansion of phase corrugations. 
With large apertures, each pupil AO should 
correct for a wide field of view, with significant gain in 
the faint source direction. 

The $\Delta$OPD model of our analysis is presented in the next Section, 
and quantities relevant to phase-referencing are recalled and clarified: 
the 2-sample variance and the attenuation factor due to fringe smearing. 
Results of numerical simulations are given in Sect.~3, with application to 
the 1.8m and 8m telescopes at Paranal. 
We further discuss the optimization of exposure duration and its dependence 
on faint star magnitude and separation angle (Sect.~4). Different regimes of 
phase-referencing are outlined in Sect.~5, together with sky coverages.

\section{Model for phase-referencing}

\subsection{Power spectra of atmospheric turbulence}
 
A detailed analysis of power spectra of quantities relevant to optical 
interferometry can be found in \cite{conan2000}, where various analytical forms 
are also proposed for integrals of power spectra. In this Section, we 
recall the main expressions.

The atmospheric turbulence, assumed to be stratified with altitude, 
is described by a distribution of discrete turbulence layers. The  
strength of each layer $k$ is characterized by an optical turbulence factor
${\cal J}_{k}$, sum of the refractive index structure parameter 
$C^{2}_{N}(h)$ along the line of sight through this layer. 
The power spectrum of atmospheric piston over a single aperture, from a 
single turbulence layer, with a von Karman turbulence spectrum and 
outer scale $L_{0}$, is:
\begin{equation}
 W_{P}({\bf f}) = 0.00969\, (f^{2}+f^{2}_{0})^{-11/6} \,{\cal J}_{k}\,F_{D}(f) 
\end{equation}
\noindent
where $f_{0}=1/L_{0}$, and $F_{D}(f)$ is the spatial filter due to finite 
aperture size, and \\ 
$F_{D}(f)\,=\,[2J_{1}(\pi D f)/\pi D f]^{2}$ for a filled circular aperture 
with diameter $D$. 

For two sources at angular distance \mbox{\boldmath $\Theta$}, 
and a turbulent layer at distance $h$ along the line of sight, the power spectrum 
of differential piston over a single aperture is obtained from the covariance 
function:

\begin{equation}
 W_{\Delta P}({\bf f}) = 2W_{P}({\bf f}) (1-\cos (2 \pi {\bf f} \cdot {\bf R})) 
\end{equation}
\noindent
with ${\bf R}=h$\mbox{\boldmath $\Theta$}.\\
\noindent
Similarly, the power spectrum of interferometric piston or OPD, with a baseline 
${\bf B}$ is:
\begin{equation}
 W_{OPD}({\bf f}) = 2 W_{P}({\bf f}) (1-\cos (2 \pi {\bf f} \cdot {\bf B})) 
\end{equation}
\noindent
and the power spectrum of differential interferometric piston ($\Delta$OPD)
is obtained with the product of the two filter functions:
\begin{equation}
 W_{\Delta OPD}({\bf f}) = 4 W_{P}({\bf f}) (1-\cos (2 \pi {\bf f} 
\cdot {\bf R})) (1-\cos (2 \pi {\bf f} \cdot {\bf B})) 
\end{equation}
With the Taylor hypothesis, turbulence is displacing as a whole with velocity 
${\bf V}_{k}$ in the turbulent layer $k$. In each layer, we take the $x$-direction 
along the wind direction (${\bf f\cdot V}_{k} = f_{x}V_{k}$). The temporal power spectrum  
of frequency $\nu$ is, for that layer:
\begin{equation}
 w_{k}(\nu) = \frac{1}{V_{k}} \int_{-\infty}^{\infty}W_{k}(\frac{\nu}{V_{k}},f_{y}) 
 df_{y} 
\end{equation}
\noindent
The temporal power spectra of independent contributions are summed together 
giving the total power spectrum ${\cal W} (\nu)$ of $\Delta$OPD, for the whole 
atmosphere crossing.

\subsection{Relevant quantities in phase-referencing}

In the following, we consider that the main OPD is tracked exactly 
on a bright star and does not contribute to interferometric noise.
The ability to follow $\Delta$OPD with time, i.e. to measure its variation 
from one sample to the next, is a critical issue in case of large phase excursion, 
i.e. for angular separation larger than some ``isopistonic'' angle. Indeed, sample 
phase measurements have to be unwrapped for long term averaging of the differential 
piston. The relevant physical quantity is not so much the magnitude of differential phase
(due to differential piston), but each sample departure from a local average.
The most useful quantity for describing 
such a process is  the 2-sample variance, or Allan variance used for the characterisation 
of frequency standards (\cite{rutman78}). The local average is simply 
made with two successive samples. Let $\tau_{0}$ be the exposure duration for each 
sample $\Delta\mathrm{OPD}_{i}$, and also the sampling interval. The 2-sample variance of the 
time series is  
$\sigma^{2}_{A}(\tau_{0})=\,\frac{1}{2}\,\langle (\Delta\mathrm{OPD}_{i+1}-
\Delta\mathrm{OPD}_{i})^2 \rangle$, 
and it can be expressed in terms of the true variance $I^{2}_{\Delta OPD}(\tau)$
by (\cite{rutman78}):  
\begin{equation}
\sigma^{2}_{A}(\tau_{0})\,=\,2[I^{2}_{\Delta OPD}(\tau_{0}) - I^{2}_{\Delta OPD}(2\tau_{0})]
\end{equation}
\noindent
with:
\begin{equation}
I^{2}_{\Delta OPD}(\tau)\,=\,2\int_{0}^{\infty}{\cal W} (\nu)
[\frac{\sin(\pi \nu \tau)}{\pi \nu \tau}]^{2} d\nu
\end{equation}

A third relevant quantity is the phase jitter within each sample,  
responsible for fringe smearing and then signal loss. 
A detailed analysis of fringe smearing in VLBI, due to 
instabilities in frequency oscillators, can be found in \cite{thompson2001}.
A similar approach applied to differential piston in dual-field optical 
interferometry yields the following expression for the attenuation factor 
due to finite exposure duration.
 
Let $\Delta \phi (t)$ be the differential interferometric phase (due to 
$\Delta$OPD), and $\tau_{0}$ the exposure duration. The (random) attenuation factor
$g_{i}$ and measured differential phase $\Delta \phi_{i}$ are given by:
\begin{equation}
g_{i} e^{j\Delta \phi_{i}} \,=\,\frac{1}{\tau_{0}} \int_{t_{i}}^{t_{i}+\tau_{0}} 
e^{j\Delta \phi (t)} dt 
\end{equation}
With a quadrature-phase scheme for fringe phase and amplitude measurements, 
the average attenuation factor useful for Signal/Noise ratio estimate, 
$G(\tau_{0})$, is taken as the square root of the quadratic average 
of $g_{i}$:
\begin{equation}
G(\tau_{0}) \,=\, \Big[\frac{1}{\tau_{0}^{2}}\int_{0}^{\tau_{0}} \int_{0}^{\tau_{0}} 
\langle \exp j[\Delta \phi (t) - \Delta \phi (t')]\rangle dt dt' \Big]^{1/2}
\end{equation}
\noindent
If $\Delta \phi$ has gaussian statistics and is a stationary process, the quantity 
between brackets is 
expressed in terms of the structure function of differential interferometric 
phase: $\sigma^{2}_{\Delta \phi}(\tau)\,=\,2\,(\frac{2\pi}{\lambda})^{2}[R(0)-R(\tau)]$,
where $\lambda$ is the optical wavelength, and $R(\tau)$ the correlation function 
or Fourier Transform of ${\cal W} (\nu)$:
\begin{equation}
R(\tau)\,=\,2\int_{0}^{\infty}{\cal W} (\nu) \cos(2 \pi \nu \tau) d\nu
\end{equation}
\noindent
Equ.~9 is now written:
\begin{equation}
G(\tau_{0})\,=\,\Big[\frac{2}{\tau_{0}} \int_{0}^{\tau_{0}} (1-\frac{t}{\tau_{0}})
\exp(-\frac{\sigma^{2}_{\Delta \phi}(t)}{2}) dt \Big]^{1/2}
\end{equation}

A similar expression is given by Buscher (\cite{buscher88}) for the smearing factor 
on main OPD, the fringe correlation function  
$\exp(-\frac{\sigma^{2}_{\Delta \phi}(t)}{2})$ being expressed in terms 
of an interferometer coherence time. 
In previous works on off-axis phase-referencing, the fringe smearing 
factor has been often approximated by 
$\exp(-2\pi^{2}\,\sigma^{2}_{res}(\tau_{0})\, /\lambda^{2})$, where 
$\sigma^{2}_{res}(\tau_{0})=2\int_{0}^{\infty}{\cal W} (\nu) [1- {\rm sinc}
(\pi \nu \tau)]^{2} d\nu$,
valid only in the weak smearing limit, that is useless for phase-referencing 
beyond the ``isopistonic'' angle.

\subsection{Turbulence model of the atmosphere}

Distribution profile of turbulence parameters above Paranal are taken 
from the results of the PARSCA campaign of March 1992 (\cite{fuchsetv93}), 
with balloon soundings and SCIDAR measurements. The average turbulence profile 
[$C_{n}^{2}(h)$] and wind amplitude profile [$V(h)$]  have been used 
by several authors for simulations of atmospheric effects
(\cite{delplan2000}; \cite{espos2000}), with a nominal seeing of 0.65''.
Continuous profiles from the PARSCA report have been re-sampled with 1 km step 
(which is about the vertical resolution of SCIDAR measurements). 
Turbulence content summed over each layer is the optical turbulence factor ${\cal J}_{k}$ 
used in the following. Wind velocity has also been averaged over each layer and 
is shown on Fig.~1. In this model, the weighted average wind velocity equals 8.5m/s.
 
\begin{figure}[ht]
\centering
\includegraphics[width=6cm,totalheight=7cm]{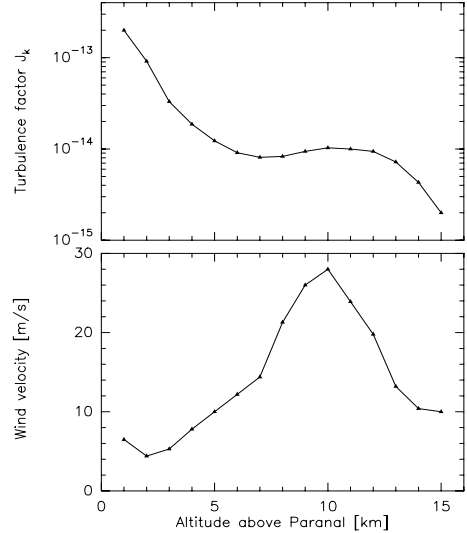}
\caption{Distribution of turbulence parameters with altitude 
above Paranal, from the average model of PARSCA 1992 campaign.
The turbulence factor is in $m^{1/3}$.
The 15 layers adopted in our model are marked with triangles.}
\end{figure}

Wind direction is another useful parameter but is  
not available in the report on the PARSCA campaign. Wind direction changes 
somehow with altitude, and scatter in wind direction smoothes
singularities of the ``frozen in'' turbulence model. We have taken a dominant wind 
direction mainly perpendicular to the interferometer baseline, that is 
a situation where turbulence effects in long baseline interferometry are 
the most significant. 

Another parameter of the model is the outer 
scale of turbulence in each one of the layers ($L_{0,k}$). This quantity 
is hardly measured locally. The Generalized Seeing Monitor (GSM) can measure 
a weighted average of $L_{0,k}$ along the line of sight. Strong turbulence layers with 
small $L_{0}$ values have a dominant effect in this weighted average (\cite{borgni90}). 
The spatial coherence outer scale measured in this way at Paranal has a log-normal 
distribution  with a median value of 22m (\cite{martin2000}) which may reflect mainly 
turbulence properties close to the telescope.  
The outer scale high in the atmosphere has a significant impact on long term averaging, 
and a limited impact on short term averaging of differential phase, as shown in the next 
Section. So, the lack of a detailed knowledge of the outer scale distribution with altitude 
does not impinge on our study.

\section{Averaging of atmospheric effects: impacts of separation angle 
and outer scale}

\begin{figure}[ht]
\centering
\includegraphics[width=7.5cm,totalheight=10cm]{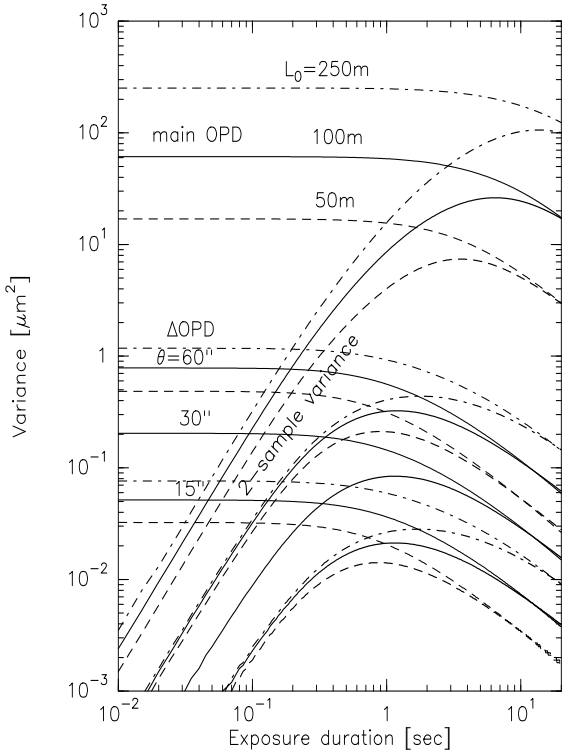}
\caption{Main OPD variance, and $\Delta$OPD variance 
for separation angles 15'', 30'' and 60'', 100m interferometer baseline and 8m 
aperture diameter. 
True variance and 2-sample variance are respectively shown with thin and thick lines. 
The other parameter is the outer scale of turbulence: 50m (dashed lines), 100m 
(full lines) and 250m (dash-point lines). For 30'' angular separation, only the full 
line is shown for clarity of the Figure.}
\label{variance} 
\end{figure}

Simulations of atmospheric turbulence have been performed with an interferometer 
baseline of 100m and telescope diameters of 8m and 1.8m. In our model, 
the separation vector between the two stars is parallel to the baseline direction, 
and the zenithal distance is nearly zero. The power spectral densities 
(main OPD and $\Delta$OPD) for each one of the 15 turbulent layers of Fig.~1 are 
estimated with Equ.~3-5, then the true variance and 2-sample variance, summed on the 
whole line of sight, are estimated with Equ.~6-7 for different exposure durations 
$\tau$. Results are shown in Fig.~\ref{variance}, both for main OPD and for $\Delta$OPD. 
Impacts of separation angles $\theta$ ranging from 15 to 60'' and of outer scale $L_{0}$ 
between 50 and 250~m are shown. 
For short exposures ($\tau \leq$ a few 0.1 s), differential phase variations, from one 
sample to the next, is small enough at optical wavelength to allow reconstruction of 
a continuous and unambiguous differential phase solution from successive samples. 
The relevant quantity for differential phase reconstruction is 
then the 2-sample variance, and not the true variance. 
This difference is essential since the true variance is a monotonic decreasing 
function with exposure duration whereas the two-sample variance increases first, 
then peaks at $\tau_{max} \simeq$ 1-2 seconds, and ultimately approaches 
the true variance for long exposures. 

In Fig.~2, the variance is found to be proportional to the square of the angular separation, 
as already shown with analytic expressions (\cite{shaocol};
 \cite{cola94}). We point also that: 
\begin{itemize}
\item the outer scale of turbulence has a significant impact both on main OPD, and on  
$\Delta$OPD for long exposures, and a small impact on the 2-sample variance 
for short exposures,
\item the variance of $\Delta$OPD remains smaller than 1$\mu$m rms, or much smaller 
than the coherence length of starlight in the near-IR (with less than 10\% relative bandwidth). 
This makes it possible differential delay tracking without feedback, that is open loop operation  
whatever the separation angle and exposure duration. 
\end{itemize}

\begin{table}[h]
\centering
\caption{Greenwood time delay $t_{0}$, isoplanatic angle $\theta_{0}$, and ``isopistonic'' 
angle $\theta_{1}$ 
for 1.8m (AT) and 8m (UT) telescopes in the $H$ and $K$ band, at zenith, at Paranal.}
\label{isoplan}
\begin{displaymath}
\begin{tabular}{l l l } 
\hline
\hline
  IR band              &    $H$     &    $K$     \\
\hline
 $t_{0}$ [ms]          &   20       &   29      \\
 $\theta_{0}$          &   8.6''   &   12.1''   \\
 $\theta_{1,AT}$      &   10.7''  &   14.3''   \\
 $\theta_{1,UT}$      &   17.4''  &   23.0''   \\
\hline
\end{tabular}
\end{displaymath}
\end{table}

The Greenwood (or AO) time delay is $t_{0}=0.314 \,r_{0}/\bar V$ (\cite{fried90}),
where $r_{0}$ is the Fried parameter and $\bar V$ the average wind velocity. 
For single aperture, the isoplanatic angle 
$\theta_{0}=[2.91 (\frac{2\pi}{\lambda})^{2} \sum_{k} {\cal J}_{k} h_{k}^{5/3}]^{-3/5}$ 
(\cite{fried82}) is defined as the angular extent over which differential wavefront 
phase corrugations is smaller than 1 radian (rms). 
By analogy, in interferometry, an angle $\theta_{1}$ can be defined as the angular range 
where $2\pi \Delta \mathrm{OPD} /\lambda$ (rms value of the true variance) is smaller 
than 1 radian. The estimation of this parameter is in Table~\ref{isoplan} for 
two different values of telescope diameter and for an outer scale of 100m. 
The $\theta_{0}$ and $\theta_{1,UT}$ estimates made by Esposito et al. (2000)
for the same observing site are somewhat smaller due to a more pessimistic turbulence 
model and to a more conservative definition of the ``isopistonic'' angle (rms phase 
of 2$\pi$/10 instead of 1 rad.). 

\begin{figure}[h]
\centering
\includegraphics[width=7cm,totalheight=6cm]{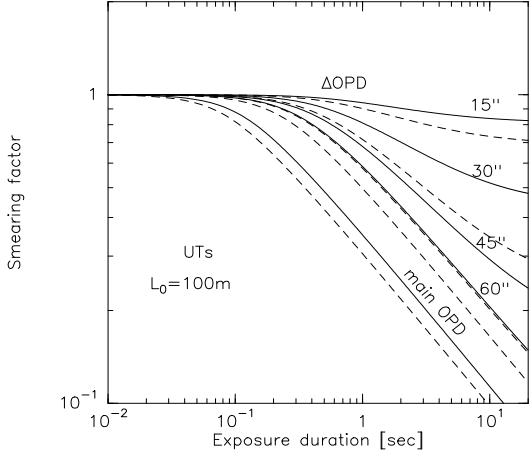}
\caption{Smearing factor due to finite exposure duration for 
differential OPD measurements with separation angles 15'', 30'', 45'' and 60'', 
in the $H$ and $K$ bands (respectively dashed and full lines). 
The smearing factor is also shown for main OPD measurements.}
\label{coher} 
\end{figure}

The average smearing factor $G(\tau_{0})$ given by Equ.~11 is shown in Fig.~\ref{coher} 
for $\Delta$OPD measurements with 4 separation angles, from 15 to 60''. This smearing factor in 
phase-referencing remains close to 1.0 for sample exposure shorter than 0.1s, for the whole 
range of angular separation. For large enough angles and long exposure duration $\tau$, 
we verify that the smearing factor is in $\tau^{-1/2}$ as shown by Buscher (\cite{buscher88}).

\section{Sensitivity estimates in phase-referencing}

Differential piston measurements is affected by the limited Signal/Noise 
ratio that can be achieved on faint sources. With $\cal B$ and $\cal S$ respectively 
the noise and signal in visibility measurement of the faint source, the measured  
2-sample differential OPD variance is the sum of two contributions, a measurement 
noise and an atmospheric differential OPD:
\begin{equation}
Y^{2}(\tau_{0})\,=\,\sigma^{2}_{m}(\tau_{0})\,+\,\sigma^{2}_{A}(\tau_{0})
\end{equation}

The measurement noise variance, given by:
\begin{equation}
\sigma^{2}_{m}(\tau_{0})\,=\, (\frac{\lambda}{2\pi})^{2}\,(\frac{{\cal B}(\tau_{0})}
{{\cal S}(\tau_{0})})^{2}
\end{equation}
\noindent
is a decreasing function of the coherent exposure duration $\tau_{0}$, whereas 
the 2-sample variance $\sigma^{2}_{A}$ of $\Delta$OPD first increases from zero to 
its peak value, and then decreases. The 
quadratic sum of the two often shows a local minimum that {\bf gives an} 
optimum exposure duration.

\subsection{Signal and Noise}

Sensitivity estimate is performed in the $H$ and $K$ bands. Measurement
noise contributions are assumed to come from read-out noise of the detector, 
and from star and background photon noise. Let $\sigma_{l}$ be the read-out noise, 
in electrons per measurement, $b_{0}$ and $s_{M}$ the number of photons 
that would be detected per second and per aperture from sky background 
and from a star with magnitude $M$, respectively, without mode filtering.
Fringe phase measurement is supposed to be performed with an ABCD scheme 
and 4 simultaneous detector readouts, also known as 'spatial discrete modulation 
scheme with 4 phases' (\cite{cassaing2000}). In the photon-rich limit, the Signal and 
Noise of such a detection scheme have been derived (\cite{dailestr99}). With 
additional mode filtering before detection, the source noise contribution is 
further reduced and we get:
\begin{equation}
{\cal S}(\tau_{0})\,=\,\gamma \Gamma G(\tau_{0}) s_{M}\tau_{0}
\end{equation}
\begin{equation}
{\cal B}^{2}(\tau_{0})\,=\,2 \sigma_{l}^{2} + ( b_{0} + \Gamma s_{M})\tau_{0}
\end{equation}
\noindent
$\gamma$ being the interferometer visibility, $\Gamma$ the average Strehl ratio 
or aperture efficiency in the point source direction, and $G(\tau_{0})$ the average fringe 
smearing factor {\bf for exposure duration $\tau_{0}$} (Equ.~11). Strehl ratio fluctuations, 
uncorrelated on the two pupils, are responsible for an additional loss of coherence 
due to unbalanced amplitudes, which is not considered here.

Parameters of SNR estimates for the UTs are given in Table~\ref{bilan}. The 
overall photometric 
efficiency is the arm transmission factor (0.30 and 0.31 in the $H$ and $K$ band respectively) 
times additional factors: coude train with two-field selection 
and AO, supposed to be 0.95 in both bands, overall efficiency of FSU (optics, 
fiber injection and mode filtering, detector quantum efficiency), supposed to be 0.3 
in each band (Cassaing, private communication). The Strehl ratio is taken from the MACAO 
document, and the interferometer visibility is an estimate. The read-out noise is taken 
as 7 electrons per measurements, with 4 frequency channels per band (spectral resolution 
$\simeq$35).
 
\begin{table}[h]
\centering
\caption{Parameters for Signal and Noise estimates with UTs.
 $E_{15}$ is the reference photon flux for a source of magnitude 15 (from \cite{cohen92},
 at Mauna Kea). 
 $s_{15}$ is the product of $E_{15}$ by the throughput (collecting area times 
 bandwidth, times overall photometric efficiency). The collecting area of each 8m 
 telescope is 49.1~m$^{2}$.}
\label{bilan} 
\begin{displaymath}
\begin{tabular}{l l l } 
\hline
\hline
IR band             		 &    $H$     &    $K$     \\
\hline
wavelength [$\mu$m]   		&   1.65     &    2.2     \\
$E_{15}$ photons/(m$^{2}$.s.$\mu$m)  &   9560   &   4560    \\
\hline
bandwidth [$\mu$m]              &  0.19        &   0.25     \\
overall photometric efficiency  &  0.085       &    0.088   \\
$s_{15}$ [photo-e$^{-}$/s]        &  7581        &   4925     \\
\hline
sky background [mag./('')$^{2}$] &   14.4       &  13.0     \\
beam size [('')$^{2}$]           &  0.01        &  0.01     \\
$b_{0}$ [photo-e$^{-}$/s]          &  132      &  310    \\
\hline
$\Gamma$ Strehl ratio            &    0.30      &  0.45      \\
$\gamma$ Interf. visibility       &   0.9        &  0.9       \\
$s_{15}$ coherent [photo-e$^{-}$/s]  & 2047      &  1995      \\
\hline
\end{tabular}
\end{displaymath}
\end{table}

A similar SNR estimate is obtained with a single frequency channel 
per band, and a read-out noise of 10 electrons instead of 7. 
Spectral resolution (e.g. for group delay measurements with a single band) 
is not so much a penalty in terms of sensitivity threshold.
Despite the smaller Strehl ratio at $H$ band,
the coherent signal from an unresolved star will be larger than at 
$K$ band. With fewer photons from sky background, phase-referencing 
at $H$ band is most useful, as shown in the next section.

\begin{figure}[h]
\centering
\includegraphics[width=6cm,totalheight=8cm]{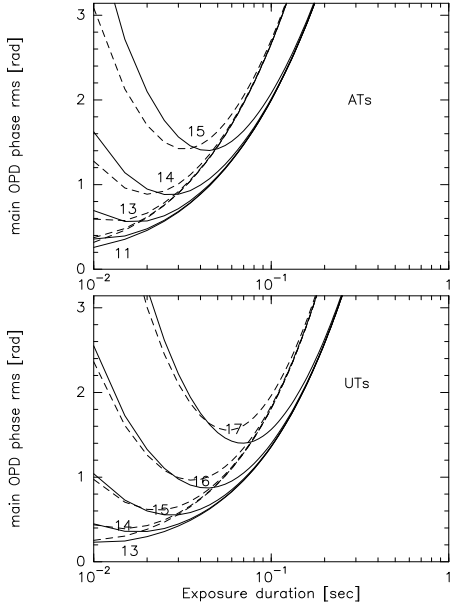}
\caption{The 2-sample phase rms of main OPD and stars with magnitude 11-15 
and 13-17 versus exposure duration  
with VLTI auxiliary telescopes (ATs) and main telescopes (UTs), in the $H$ and $K$ bands 
(respectively dashed and full lines). The outer scale of turbulence is $L_{0}$=100m.
The sensivity gain with UTs is about 2 magnitudes compared to ATs.}
\label{mainOPD} 
\end{figure}

Sensitivity estimates have been performed also with ATs. Starting from the 
UT's parameters (Table 2), collecting area is divided by 20, 
beam size is multiplied by 10, and Strehl ratios are taken as 0.54 and 0.72 in 
the $H$  and $K$ bands respectively (Tip-Tilt corrections only). The other parameters 
(spectral resolution, read-out noise \ldots) are kept unchanged.

\subsection{Magnitude limits}

The 2-sample phase rms reached with piston and measurement noise is:

\begin{equation}
y(\tau_{0})\,=\,\frac{2\pi}{\lambda}\,Y(\tau_{0})
\end{equation}

Main OPD phase is first plotted in Fig.~\ref{mainOPD}, for star magnitudes ranging from
11 to 17. A minimum rms value is obtained 
for an exposure duration $\tau_{m}$. The exposure $\tau_{opt}$ which optimizes 
the SNR per time unit will be slightly smaller than $\tau_{m}$. The minimum phase rms 
is about the same in the two bands, and stars as faint as $H$ and/or $K$ = 15 might 
be used for main delay tracking with 8m telescopes. Close loop operation will degrade 
such a figure of merit, so that the magnitude limit will most probably lie in the 
range 13-14.

\begin{figure}[h]
\centering
\includegraphics[width=6cm,totalheight=10.5cm]{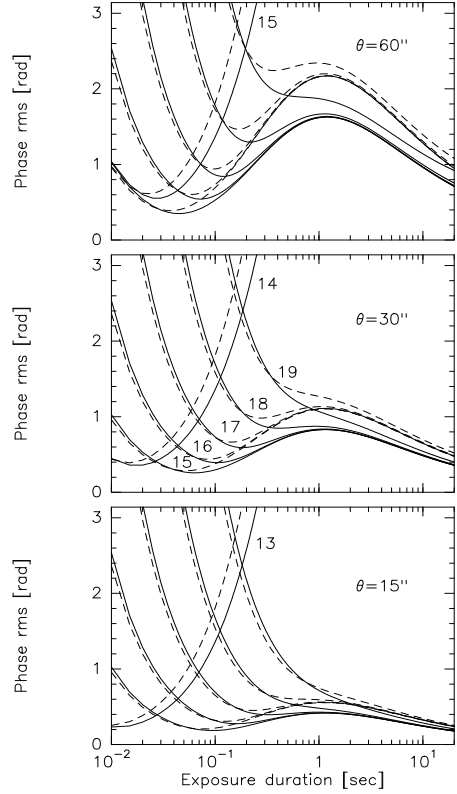}
\caption{The 2-sample phase rms of main OPD
(thick lines) and $\Delta$OPD (thin lines) in the $H$ and $K$ bands 
(respectively dashed and full lines), with 8m telescopes and $L_{0}=100$m,
for three values of separation angle $\theta$. 
On each graph, the magnitude range for $\Delta$OPD is from 15 to 19. 
A single magnitude is shown for main OPD.}
\label{DeltaOPD} 
\end{figure}

The 2-sample differential phase rms measured with the UTs is plotted in 
Fig.~\ref{DeltaOPD}, 
for separation angles 15'', 30'' and 60'', and for faint star magnitude 
in the range 15 to 19.
Again a minimum is reached for stars not too faint. The exposure time 
at minimum increases with magnitude, and slightly decreases with 
separation angle. The range we find for an optimum exposure is from 
0.03 to 0.5 seconds.

\section{Phase-referencing with short exposure}

 \begin{figure}[h]
\centering
\includegraphics[totalheight=6cm]{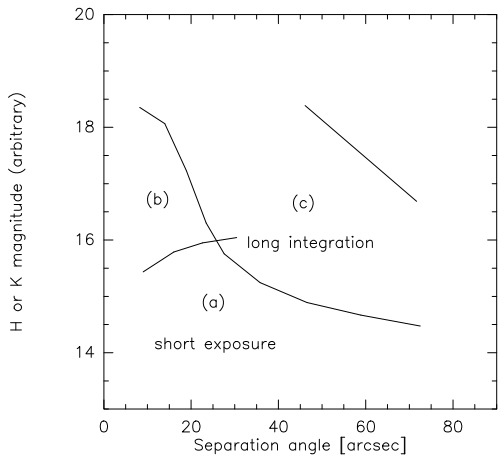}
\caption{Different regimes of Phase-Referencing based on the 2-sample differential phase. 
In the (a) domain, there exists a well defined minimum satisfying $\Delta$OPD phase rms 
smaller than 1 rad for an optimum exposure duration $\tau_{opt} \leq$ 1s. Domain (b) 
corresponds to short exposure without a well defined optimum exposure duration.
Domain (c) corresponds to long integration, limited in the upper right corner 
by a straight line with a somewhat arbitrary threshold on phase rms, taken here to be 
smaller than 1 rad for 20 seconds integration. }
\label{PR-zones} 
\end{figure}

The minimum phase variation reached for $\tau=\tau_{m}$ is about the same in the 
$H$ and $K$ bands, so that we should consider the benefit 
of two-band observing. Let $\sigma_{0}$ be the minimum phase variation (rms) 
at the effective wavelengths $\lambda_{H}$ and $\lambda_{K}$.
At minimum, the contributions of differential piston and measurement 
noise to the measured 2-sample variance are about the same (Equ.~12).
Measurement noises in the two bands being uncorrelated, 
the uncertainty on $\Delta$OPD variation, from one sample to the next, is:
\begin{equation}
\sigma_{\delta l}\,\simeq \,\frac{\sigma_{0}}{4 \pi}(\lambda_{H}^{2}\,+\,\lambda_{K}^{2})^{1/2}
\end{equation}
For $\sigma_{0}$=1 radian, the uncertainty on $\Delta$OPD variation is about 
0.21~$\mu$m that is less than $\lambda/10$ in the $K$ band.
Such a figure is obtained with an equivalent SNR of about 1.6.
Unwrapping differential phase measurements, from one sample to the next, should be 
possible with filtering algorithms reducing the contribution of measurement noise. 
Indeed, in the considered frequency range of about 10 Hz, the $\Delta$OPD 
power spectrum will have a steep slope due to spatial filtering with the 
telescope aperture, whereas measurement noise will have a flat spectrum. 
Its contribution could be reduced with some time averaging.

A domain of SPR with short exposure is defined in a diagram
(separation angle, magnitude of the faint object), with the condition:
``there exists a minimum in the 2-sample $\Delta$OPD phase, and 
the minimum value is smaller than 1 radian (rms)''. An optimum exposure duration 
for faint source observing is then clearly defined. This domain is shown with 
label (a) in the lower part of Fig.~\ref{PR-zones}. For fainter sources and separation 
angle not too large ($\leq$~30''), there may be no minimum, but a measured 2-sample 
phase rms smaller than 1 radian for exposures smaller 
than about 1 second, as, for example, $K \geq 18$ with $\theta = 15''$ (Fig.~5).
Phase-referencing is still possible with short exposure, 
but without an optimum duration. Uncertainty is then dominated by measurement 
noise, and this domain is labelled (b) in Fig.~\ref{PR-zones}. The third domain, 
labelled (c), is the usual domain for long exposures. The exposure 
duration should be long enough in order to smooth out fluctuations, that is well  
beyond the peak of the 2-sample variance.  
As shown in Fig.~\ref{DeltaOPD}, phase-referencing of a 19th magnitude source is 
marginally possible up to a separation angle of 1 arcminute, the phase 
rms being about 1 radian for 20 seconds integration.

\begin{figure}[h]
\centering
\includegraphics[totalheight=7cm]{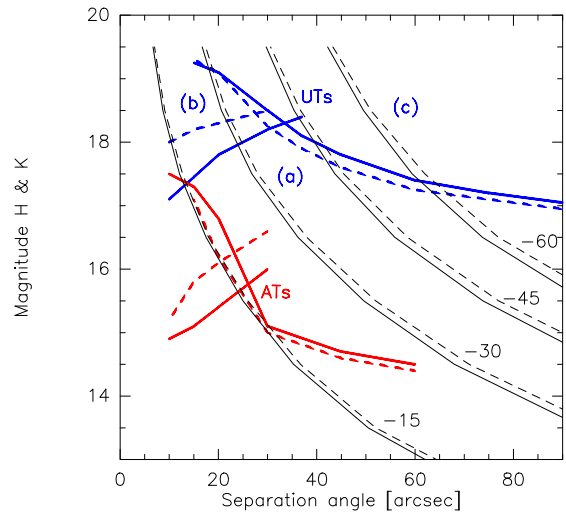}
\caption{Accessibility to faint stars in terms of magnitude 
and separation angle in the $H$ and $K$ bands (respectively dashed and full lines).
The 3 domains of Fig.~6 for SPR are recalled here with thick lines for 
ATs and UTs observations. The upper right limit of the domain (c) is omitted for clarity.
The thin curves indicate that there are 3 stars in average within the half cone of the 
corresponding separation and brighter than the corresponding magnitude. These curves 
are labelled with the galactic latitude of the target direction.}
\label{PR-stars} 
\end{figure}

The newly identified domain (a) for SPR with short exposure is useful 
only if there are stars satisfying the magnitude and separation constraints shown
in Fig.~\ref{PR-zones}. We first consider phase-referencing for the 
astrometric search of exoplanets orbiting bright stars, i.e. bright enough for 
main delay tracking. We would like to measure the angular separation between 
a main target and reference stars in its surrounding. As randomly choosen 
stars can turn out to be binaries, we take a conservative value 
of 3 reference stars (in average) within the useful solid angle. Reference 
stars for differential astrometry will most often be faint objects, so that we 
must rely on star count models for their distribution with galactic coordinates.
Our results are based on the 
Besan\c{c}on model on stellar populations in our Galaxy (\cite{reylerobin01}).    
The half cone angle with 3 stars (in average) brighter than a given magnitude $m$ 
is taken as $[N(m)]^{-1/2}$, where $N(m)$ is the cumulative density of stars 
brighter than $m$. Results shown in Fig.~\ref{PR-stars} are for a region of the 
sky suited for phase measurements, with a nearly zenith transit at Paranal. 
Its galactic longitude is $-15$ degrees, and the latitude range is from $-15$ 
to $-60$ degrees. Whatever the target direction, 3 stars are likely to be found 
with UTs and SPR with short exposure (domain (a)). At a galactic latitude 
of $-60$ degrees, the separation angle reaches 1 arcmin and faint star 
magnitude 17.5. Long exposure SPR with fainter and closer stars is also 
possible, say 19.5 magnitude and about 42'', but in a regime where the useful 
photon flux from the star is one tenth the photon flux from the sky background in 
the $K$ band. Closer to the galactic plane, SPR with short exposure is much easier 
due to the greater star density. 
\begin{figure}[h]
\centering
\includegraphics[width=6.5cm,totalheight=11cm]{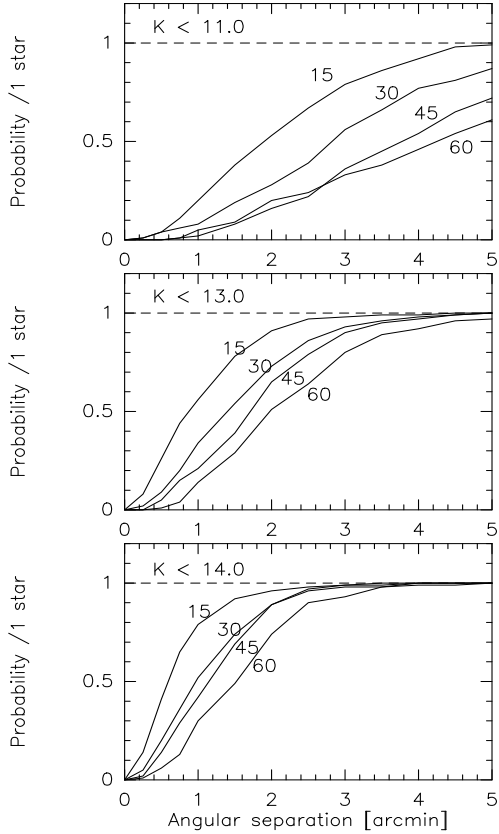}
\caption{Probability of finding a guide star for main delay tracking 
in the vicinity of any target direction with galactic latitudes 
15, 30, 45 and 60 degrees.
The guide star has to be brighter than K=11.0 (top), 13.0 (middle) 
or 14.0 (bottom).}
\label{probtot} 
\end{figure}

A different estimate has to be performed for phase-referenced imaging of a faint 
object. What is needed is a bright star in its surrounding for main delay tracking. The 
bright (or guide) star lies in the magnitude range of near-IR surveys, and 
our estimate of sky-coverage is based on the Second Release of the 2MASS survey
\footnote{This publication 
makes use of data products from the the Two Micron All Sky Survey, 
which is a joint project of the University of Massachusetts and the Infrared 
Processing and Analysis Center/California Institute of Technology, funded by the 
National Aeronautics and Space Administration and the National Science Foundation.}. 
The probability of finding at least one star brighter than $K_{lim}$ has been 
estimated in terms of the size of the solid angle for phase-referencing, at different 
galactic latitudes (Fig.~\ref{probtot}). Results on positive and negative galactic 
latitudes have been averaged, and the magnitude limit for the guide star is 
$K_{lim}$ = 11 (with two ATs), and 13 or 14 (with two UTs). Taking a galactic latitude 
of 30 degrees as an average sky density, the probability of finding a guide star 
with 2 ATs ($K_{lim} \leq 11$, within 20'') is about 2\%. 
Significant sky coverage, with a probability of 50\%, will be reached only with 2 UTs 
and improved corrections for the atmospheric turbulence ($K_{lim} \leq 14$ 
and AO corrections in two different directions, up to 1' apart).

\section{Discussion}

The useful angular range for phase-referencing and differential astrometry is not strictly 
limited, but dependent on the magnitude of the faint object  (Fig.~\ref{PR-stars}). 
Then, the `isopistonic' angle that has been defined in the literature is too restrictive.
In our analysis, there exists a specific angle given by the `triple-point' in 
Fig.~\ref{PR-zones}, where the 2-sample phase rms equals 1 radian at a local  
minimum. The separation angle and the magnitude of the faint 
object at this triple-point depend both on the interferometer sentivity and on the 
site turbulence. It characterizes a whole system (atmosphere + telescope + instrument). 
An interesting finding of our calculation is that the triple-point is located at about 
(20'', 16) for ATs and (30'', 18.3) for UTs with similar instruments. This makes a 
very significant difference in sky coverage, differential astrometry with ATs 
being limited to rather dense stellar fields (Fig.~\ref{PR-stars}). 

Short exposure measurements of differential phase fluctuations integrate 
various contributions: differential piston, sky background variations, 
interferometer/instrument instabilities \ldots 
Short term measurements can then be edited, filtered and weighted 
before being averaged to properly extract source structure and relative position information. 
The similar sensitivities and star counts obtained in the $H$ and $K$ bands 
(Fig.~\ref{PR-stars}) must be stressed, as dual-band observations allow more robust
procedures than single band near the magnitude limit.

In our analysis, Adaptive Optics has not been addressed, whereas wavefront 
corrections are needed in the two sources direction, at least with large telescopes. 
Tomographic sensing of the atmospheric turbulence with the observation of several 
guide stars has been proposed for wavefront corrections on a whole telescope 
FOV (\cite{tallon1990}). The technique is known as Multi Conjugate Adaptive Optics 
(MCAO). Although mainly concerned with improved Strehl Ratio, MCAO is supposed 
to correct for atmospheric turbulence in a whole volume, and hence to correct 
for differential pathlength fluctuations on any entrance pupil. Differential 
interferometric piston ($\Delta$OPD) should be significantly reduced, allowing for 
long term averaging in faint object imaging on a whole telescope FOV. However,
it is difficult to know how much $\Delta$OPD would be reduced since corrugations 
with large scale lengths, about the pupil size, are less efficiently corrected 
than shorter scale lengths.
 
Sky coverage with natural guide stars was thought to be a main problem for AO  
and not too large telescopes.
An efficient technique for MCAO has been proposed recently (\cite{raga2002}).
It is layer-oriented, with multiple Field of View: a large annular FOV is used 
for sensing low altitude turbulence, whereas a narrower FOV (about 2') in the 
telescope axis is used for sensing high altitude turbulence. Sky coverage 
exceeding 50\% should be reached with 8m telescopes and a wavefront sensor 
in the R band.
Such a sky coverage is quite consistent with our estimate on spatial phase-referencing
in dual-field optical interferometry in the near-IR.     
It makes faint source imaging with VLTI/UT a very promissing next development step.

With the AO system being presently implemented at the Coud\'e focus of UTs (MACAO), 
faint source imaging will be restricted to the isoplanatic patch of guide stars. 
It may reach the telescope FOV of 2 arcmin when turbulence in the upper layers of the 
atmosphere is particularly weak. 
For bright enough targets, our analysis shows that $\Delta$OPD will be measured 
and tracked well beyond the isoplanatic angle.
 
\begin{acknowledgements}
We acknowledge helpful discussions with F. Cassaing and J.-M. Conan. We are 
grateful to A. Robin for providing us with near-IR star counts from the 
Besan\c{c}on Galaxy model. We greatly acknowledge A. Quirrenbach 
for very helpful comments and suggestions in his refeering of the paper.  

\end{acknowledgements}

\end{document}